\documentclass[12pt,preprint]{aastex}

\usepackage{amsmath}

\usepackage{color}
\usepackage[normalem]{ulem}

\slugcomment{accepted in The Astrophysical Journal}

\shorttitle{A Rapid Water Loss can Extend the Lifetime of the Planetary Habitability}
\shortauthors{Kodama et al.}

\begin{document}


\title{Rapid Water Loss can Extend the Lifetime \\ of the Planetary Habitability}


\author{Takanori Kodama}
\affil{Department of Earth and Planetary Science, The University of Tokyo, Tokyo 113-0033, Japan.}
\email{koda@eps.s.u-tokyo.ac.jp}

\author{Hidenori Genda}
\affil{Earth-Life Science Institute, Tokyo Institute of Technology, Tokyo 152-8550, Japan.}

\author{Yutaka Abe}
\affil{Department of Earth and Planetary Science, The University of Tokyo, Tokyo 113-0033, Japan.}

\and

\author{Kevin J. Zahnle}
\affil{Space Science and Astrobiology Division, NASA Ames Research Center, California 94035, USA.}


%%\altaffiltext{1}{Visiting Astronomer, Cerro Tololo Inter-American Observatory.
%%CTIO is operated by AURA, Inc.\ under contract to the National Science
%%Foundation.}
%%\altaffiltext{2}{Society of Fellows, Harvard University.}
%%\altaffiltext{3}{present address: Center for Astrophysics,
%%    60 Garden Street, Cambridge, MA 02138}
%%\altaffiltext{4}{Visiting Programmer, Space Telescope Science Institute}
%%\altaffiltext{5}{Patron, Alonso's Bar and Grill}


\begin{abstract}

Two habitable planetary states are proposed: an aqua planet like the Earth and a land planet that has a small amount of water. Land planets keep liquid water under larger solar radiation compared to aqua planets. Water loss may change an aqua planet into a land planet, and the planet can remain habitable for a longer time than had it stayed an aqua planet. We calculate planetary evolution with hydrogen escape for different initial water inventories and different distances from the central star. We find that there are two conditions to evolve an aqua planet into a land planet: the critical amount of water on the surface ($M_\mathrm{ml}$) consistent with a planet being a land planet, and the critical amount of water vapor in the atmosphere ($M_\mathrm{cv}$) that defines the onset of the runaway greenhouse state. We find that Earth-size aqua planets with initial oceans $< 10 \%$ of the Earth's can evolve into land planets if $M_\mathrm{cv} = 3 \, \mathrm{m}$ in precipitable water and $M_\mathrm{ml} = 5 \, \%$ of the Earth's ocean mass. Such planets can keep liquid water on their surface for another $2$ Gyrs. The initial amount of water and $M_\mathrm{cv}$ are shown to be important dividing parameters of the planetary evolution path. Our results indicate that massive hydrogen escape could give a fresh start as another kind of habitable planet rather than the end of its habitability.

\end{abstract}





%*************************************** Introduction **********************************************

\section{Introduction}

	The advancement of observation technologies has facilitated an increase in the number of extrasolar planets detected. Some of these planets are considered to be terrestrial or rocky planets based on the relationship between their mass and radius \citep[e.g.,][] {Bat13}. Although it is not clear how much water exists on these planets, the marked variations in their bulk densities imply that considerable variation exists in the amount of water on these extrasolar terrestrial planets. The habitability of extrasolar planets has been actively discussed \citep[e.g.,][] {Bor13}. In particular, the stability of liquid water on the planetary surface has been extensively examined as liquid water is considered to be essential for the origin and evolution of life.

	It is typical to assume that habitable planets, like the Earth, have liquid water oceans. Discussions of planetary habitability and the duration of planetary habitability (known as continuous habitability) have focused on the stability and evolution of such oceans. If the insolation that a planet covered with an ocean receives from a central star is not too high, then the upper atmosphere of such a planet would have a very low mixing ratio of water vapor due to the presence of a cold trap. Under these conditions hydrogen escape is rather slow, and partly dependent on the presence of other gases containing hydrogen (e.g. methane). As the luminosity of the central star increases with time, the planet gets warmer and the mixing ratio of water vapor in the upper atmosphere increases. When the central star gets bright enough, the cold trap disappears and the mixing ratio of water vapor in the upper atmosphere becomes high enough to cause rapid water loss. This planetary state is called the moist greenhouse state \citep{Kas93}. Somewhat later, the star becomes bright enough to trigger the runaway greenhouse effect: because water vapor is a strong greenhouse gas, a positive feedback works until all liquid water evaporates in the runaway greenhouse state \citep[e.g.,][] {Abe88, Kas88, Zah88, Nak92}.

	The habitable zone (HZ) has been defined as the region around the central star where liquid water is stable on the planetary surface \citep[e.g.,][] {Kas93, Kop13}.  The HZs for various planetary masses around central stars with various spectral types are also investigated \citep[e.g.,][]{Kop14}. Two concepts have been proposed to describe the inner edge of the habitable zone. One is to define it as the inner edge of the instantaneous habitable zone; that it is, it does not matter how long the planet remains habitable. The other defines it as the inner edge of the continuously habitable zone (CHZ), by which it is meant that a given planet has been in the habitable zone from the time it was formed until today. The onset of the runaway greenhouse effect determines the instantaneous habitable zone. When considering the long-term stability of liquid water on the planetary surface, the water loss from the planet is an important consideration. Under moist greenhouse state, hydrogen escapes rapidly. If the escape of hydrogen occurs rapid enough, then the inner edge of the CHZ may be characterized by the onset of the moist greenhouse state \citep[e.g.,][] {Kas93, Kop13}.

	It has recently been shown that the climates of planets with small or large amounts of water will differ markedly from each other \citep[]{Abe05, Abe11}. Using a general circulation model (GCM), \citet{Abe11} examined the habitable zones of land planets. They found that the local balance between the precipitation and evaporation of water is controlled by the distribution of liquid water on the planetary surface. Liquid water typically accumulates in the cooler regions of a planet, that is, high-latitude regions and low-latitude regions typically become dry. Given that there is no upper limit on planetary radiation in dry areas \citep[e.g.,][] {Nak92}, a land planet can maintain liquid water on its surface at much larger incident of stellar radiation than is possible on aqua planets. In addition, the mixing ratio of water vapor in the upper atmosphere is very low on land planets, which means that the hydrogen escape flux should also be very low. However, the liquid water on a land planet will evaporate if the incident stellar radiation is high enough. This radiation is called the threshold of the runaway greenhouse for a land planet. According to \citet{Abe11}, this is $415 \, \mathrm{W/m}^2$. The inner edge of the habitable zone for a land planet is determined by this threshold, which corresponds to $0.77 \, \mathrm{AU}$ in our solar system. For aqua planets, the onset of the runaway greenhouse effect is about $375 \, \mathrm{W/m}^2$ \citep{Lec13}. Therefore, the inner edge of the habitable zone for a land planet is much closer to the central star than that for an aqua planet.

	The only difference between a land planet and an aqua planet is the amount of water that the planets have. If an aqua planet evolves to a land planet by rapid water loss before all of the water on its surface evaporates, then the lifetime of liquid water on its surface can be extended because land planets are highly resistant both to water loss and to complete evaporation. In this way, rapid water loss might lead to maintaining, rather than ending, the habitable world.

	When \citet{Abe11} discussed the evolution from an aqua planet to a land planet, they assumed that the aqua planet can evolve to the land planet when the planetary surface becomes almost completely dry before the onset of the runaway greenhouse made the planet uninhabitable. They estimated how much water could escape from a planet in Earth's orbit using models that assumed both constant ($0.3$) and variable albedos. For the constant albedo scenario, an aqua planet with Earth's ocean mass did not lose all of its water before the onset of the runaway greenhouse, because the duration of the moist greenhouse state was too short. Such planets lapsed into the runaway greenhouse state and became uninhabitable. Conversely, in the variable albedo scenario, the increase of the albedo due to Rayleigh scattering prolonged the duration of the moist greenhouse state. In this case, the aqua planet with Earth's ocean mass lost almost all of its surface water before the onset of the runaway greenhouse state, and then evolved into a land planet. 
Here, we investigate the evolution from the aqua planet mode to the land planet mode invoking more realistic transition conditions for the amount of water vapor in the atmosphere and the amount of water on the planetary surface.

In the evolution from the aqua planet to the land planet, there is a race between rapid water loss and the increase in the luminosity from the central star that triggers the runaway greenhouse effect. Whether or not an aqua planet safely negotiates the evolutionary path from the aqua planet mode to the land planet mode is determined by the outcome of this race. \citet{Abe11} only considered the evolution from aqua planet mode to land planet mode for planets that have one Earth ocean mass in either the Earth's orbit or Venus's orbit. However, discoveries of many extrasolar planets have revealed the existence of a great diversity of planet types, and we consider that the amount of water that planets have should reflect this diversity.

We therefore systematically investigated evolutionary scenarios in which aqua planets can evolve into land planets under conditions with different initial amounts of water and distances from the central star. We discuss the evolution from aqua planet mode to land planet mode and re-evaluate the inner edge of the HZ. In addition, we investigate the effect of varying the transition conditions that determine whether an aqua planet can evolve into a land planet.

%*************************************** Introduction **********************************************

%*************************************** Evolution model **********************************************

\section{Evolution model}

%% In a manner similar to \objectname authors can provide links to dataset
%% hosted at participating data centers via the \dataset{} command.  The
%% second curly bracket argument is printed in the text while the first
%% parentheses argument serves as the valid data set identifier.  Large
%% lists of data set are best provided in a table (see Table 3 for an example).
%% Valid data set identifiers should be obtained from the data center that
%% is currently hosting the data.
%%
%% Note that AASTeX interprets everything between the curly braces in the 
%% macro as regular text, so any special characters, e.g. "#" or "_," must be 
%% preceded by a backslash. Otherwise, you will get a LaTeX error when you 
%% compile your manuscript.  Special characters do not 
%% need to be escaped in the optional, square-bracket argument.

In this paper, we consider three planetary states (a water planet state, a steam planet state, and a dry planet state), and two modes for a water planet state (an aqua planet mode and a land planet mode). Here, we describe the main characteristics of these states and modes in advance.

In a water planet state, liquid water is stably present on the planetary surface. In most previous studies, planets in the aqua planet mode are considered to be potentially habitable \citep[e.g.,][] {Kas93, Kop13}. In these studies, it has been implicitly assumed that planets have a large amount of water (e.g., the present Earth's oceans). On the other hand, if planets have a small amount of water, then they are considered to be in land planet mode \citep[]{Abe11}.

Planets in a steam planet state have water vapor in their atmosphere but no liquid water on their surface. If planets with liquid water receive the radiation exceeding the threshold for the runaway greenhouse effect, then all of the liquid water on their surface will evaporate and these planets will evolve into a steam planet state.

In a dry planet state, planets have no liquid water on their surface and no water vapor in their atmosphere. Dry planets are not the same as land planets, as the latter have a small amount of liquid water on their surface. If water vapor escapes from a steam planet, then it becomes a dry planet and both steam planets and dry planets are considered to be uninhabitable.

Here we simulate the changes of the amount of water on the planetary surface and in the atmosphere using a numerical model. In section 2.1, we describe the basic processes that cause changes in the amount of atmospheric and surface water. The transition conditions from aqua planet mode to land planet mode are described in Section 2.2. In Section 2.3, we describe the hydrodynamic escape of hydrogen as a rapid water loss. The escape flux of hydrogen molecules depends on the mixing ratio of hydrogen molecules in the upper atmosphere and the evolution of the luminosity and EUV flux of the central star, which are described in sections 2.4 and 2.5, respectively.

\subsection{The processes that change the amount of water}

The initial amount of water on terrestrial planets just after their formation is not known. We treat the initial amount of water as a parameter, i.e., we consider planets with various initial amounts of water. The amount of water on the surface of a planet would change through the interaction with the planetary interior through processes such as subduction of hydrous minerals, outgassing of water at mid-ocean ridges and arcs, and so on. For the present Earth, the ingassing flux of water is estimated to be $ 5 \times 10^{22} - 1.1 \times 10^{23} \, \mathrm{mol/Gyr}$  \citep{Rea96, Jov98}, and the degassing flux of water is estimated to be $ 2.2 \times 10^{22} - 1.1 \times 10^{23} \, \mathrm{mol/Gyr}$  \citep{Bou01, Hil02}. Ingassing and outgassing fluxes in water appear to be in equilibrium on the present Earth. The difference between the ingassing and outgassing water fluxes is the order of $10^{22} \, \mathrm{mol/Gyr}$. By comparison, once the rapid escape of water into space starts as the central star evolves, the escape flux of water vapor exceeds $10^{23} \mathrm{mol/Gyr} $(see Sections 3.1 and 3.2). For simplicity, we assume that the change in the total amount of water on a planet is attributed solely to the loss of water into the space.

We assume that the amount of water decreases via the hydrodynamic escape of hydrogen molecules. The change in the amount of water is given by
\begin{equation}
  \frac{dM_\mathrm{H_{2}O} }{dt} = - \phi_\mathrm{esc}
\end{equation}
where $M_\mathrm{H_{2}O}$ is the total amount of water and $\phi_\mathrm{esc}$ is the escape flux of molecular hydrogen. The escape process of hydrogen is described in Section 2.3.

The amount of water vapor in the atmosphere depends on the atmosphere model used. For the aqua planet and land planet modes, the amount of water vapor is estimated by using the $1$-D, cloud-free, radiative-convective atmospheric model developed by \citet{Abe88} and using the results obtained by \citet{Abe11}, respectively. The amount of water on the surface can then be estimated by subtracting the amount of water in the atmosphere from the total amount of water on the planet. An aqua planet is considered to evolve into a land planet if the amounts of water in both reservoirs meet specific transition conditions, described in detail in Section 2.2.

Planetary albedo, which plays an important role in climate, depends on the properties of the planetary surface and atmosphere. In our numerical simulation, we fixed the albedo of the planetary surface at $0.3$, which is typical of deserts. When the amount of water vapor increases due to warming caused by an increase in the luminosity of the central star, the atmosphere gets denser and light from the central star is scattered more efficiently. Therefore, we also consider the scattering albedo for an aqua planet. Since the scattering albedo is negligible for a land planet (if the effect of dust is not considered), we consider that the planetary albedo for a land planet to be $0.3$.

Here, we assume that the planets have zero obliquity and eccentricity, with a mass and a radius equivalent to those of the Earth. In the planetary atmosphere, $1$ bar of the present Earth's atmosphere is considered as the background atmosphere. Except for water vapor, no changes in atmospheric components were considered in this study. It is possible that the carbon cycle exists on Earth-like planets. The carbon cycle in the presence of liquid water provides a negative feedback on surface temperature that maintains a stable climate \citep[]{Wal81}. However, in the case of a central star with high luminosity, the feedback breaks down because there is too little CO$_2$ in the atmosphere to affect changes in surface temperatures owing to efficient chemical erosion (rapid fixing of CO$_2$ as carbonate rock). Therefore, the evolution of the atmosphere is driven almost entirely by changes in the amount of water in the planetary atmosphere.

\subsection{Transition conditions}

\citet{Abe11} assumed that an aqua planet with the present Earth's ocean mass could evolve to land planet mode just before all of its water is lost, and before there was a rapid increase in surface temperatures due to the runaway greenhouse effect. Under such transition conditions, it is possible that a planet with a large amount of water in its atmosphere can evolve from an aqua planet to a land planet. However, if the planet has a large amount of water vapor in its atmosphere before the loss of water, then it may maintain a heated state during atmospheric escape. Under these conditions, such a planet may not evolve into a land planet, even if it loses most of the water on the surface. It is therefore necessary to consider the amount of water vapor in the planetary atmosphere. We therefore imposed several conditions on amounts of water on the planetary surface and amounts of water vapor in its atmosphere for the transition from an aqua planet to land planet mode.

However, one problem is that we do not know how much water vapor there is in an aqua planet's atmosphere just before the threshold for the runaway greenhouse effect is reached. In order to evolve to land planet mode, there should be an upper limit to the amount of water in the atmosphere; we treat this upper limit to the amount of water as a parameter, which we denote as $M_\mathrm{cv}$  and refer to as "the critical amount of vapor for the runaway greenhouse effect". From previous studies \citep[]{Abe88, Kas88, Kop13, Gol13}, it appears that when $M_\mathrm{cv}$  is expressed as a column it is equivalent to $1 \, \mathrm{m}$ to $10 \, \mathrm{m}$ of precipitable water. In this paper, to illustrate typical results, an $M_\mathrm{cv}$  of $3 \, \mathrm{m}$ was used for our standard model.

Another important parameter is the maximum amount of water that a land planet can have and still behave as a land planet. We consider the distribution of surface water from the results of \citet{Abe11}. Surface water on a land planet is typically located at latitudes above $60$ degrees, which corresponds to $10 \,\%$ of the entire planetary surface. According to percolation theory, for any given area, only half of that area is required to form a connected area. The maximum unconnected oceanic area is thus half of the area above $60$ degrees latitude, which corresponds to $5\%$ of the entire planetary surface. If planetary topography is ignored, then the typical amount of water on a land planet is $5 \,\%$ of the present Earth's ocean mass. If a planet has more water, that water will extend equatorward of 60 degrees latitude. Therefore, we assume that $5 \, \%$ of the present Earth's ocean mass on the planetary surface approximates the transition conditions that are required for the maximum amount of water required for an aqua planet to evolve into a land planet. We refer to this condition ``the maximum liquid water mass for a land planet mode'', which we denote as $M_\mathrm{ml}$. If the amounts of water on the planetary surface and in the atmosphere meet these conditions, then the aqua planet can evolve into a land planet. We treat $M_\mathrm{ml} = 0.05 \, M_\mathrm{oc}$ as our standard case, but we also treat $M_\mathrm{ml}$ as a parameter that varies from $0.01$ to $0.1 \, M_\mathrm{oc}$, where $M_\mathrm{oc}$ is the amount of the present Earth's ocean ($8.4\times10^{22} \,  \mathrm{moles} $). In Section 3.5, we discuss the evolution of an aqua planet under several transition conditions.

\subsection{Water loss}

Escape is an important process in atmospheric evolution. In this study, we consider the hydrodynamic escape of hydrogen molecules because it is the fastest process of hydrogen escape and has the most potential to decrease the water reservoir of a planet. We will address two major bottlenecks to hydrogen molecular escape from the planetary atmosphere. The first is the energy source required to drive the escape of hydrogen molecules. The main source of energy for escape is extreme ultraviolet radiation (EUV), usually defined by $\lambda <100 \, \mathrm{nm}$ from the central star, which is directly absorbed by hydrogen molecules and atoms. When the escape flux is controlled by the EUV flux, the escape is said to be ``energy-limited''. The other major bottleneck is the diffusion of hydrogen-containing molecules through the background atmosphere to reach the altitudes where escape occurs. When escape is controlled by the diffusion of molecules through the upper atmosphere, then the escape is referred to as ``diffusion-limited''. We calculate escape fluxes in both modes and adopted the smaller one as the actual escape flux. Other possible bottlenecks, such as the photon-limited photochemistry that converts H$_2$O to H$_2$, magnetospheric drag, and radiative cooling by ions (e.g., H$_3^+$) and molecules embedded in the outflowing wind, are less well established and will be neglected here.

	The escape flux of the diffusion-limited escape mode depends on the total mixing ratio of hydrogen-containing species in the upper atmosphere \citep[]{Wal77}. The diffusion-limited escape flux ($\phi_\mathrm{d}$ ) of hydrogen molecules is given by
\begin{equation}
  \phi_\mathrm{d} = f_\mathrm{T} ( \mathrm{H}_2 ) \frac{b ( m_\mathrm{a} - m_\mathrm{i}) g }{k T_\mathrm{str}}
\end{equation}
where $f_\mathrm{T} ( \mathrm{H}_2 )$ is the total mixing ratio of hydrogen molecules in all forms at homopause
$( f_\mathrm{T} ( \mathrm{H}_2 ) =f( \mathrm{H}_2 \mathrm{O} ) + f( \mathrm{H}_2 ) + 2 f ( \mathrm{CH}_4 ) + \cdots)$, 
$T_\mathrm{str}$ is the temperature in the stratosphere ($T_\mathrm{str} = 200 \, \mathrm{K}$),
$b$ is the binary diffusion coefficient between the background atmosphere and $\mathrm{H}_2$
( $b = 1.9 \times 10^{19} ( T_\mathrm{str} / 300 \, \mathrm{K})^{0.75} \, \mathrm{cm}^{-1} \mathrm{s}^{-1}$ for $\mathrm{H}_2$ in air),
 $g$ is the acceleration due to gravity, $k$ is Boltzmannâs constant, $m_\mathrm{a}$ and $m_\mathrm{i}$ are the average molecular masses of the background atmosphere and the molecular mass of hydrogen, respectively. 
 We assumed that the major carrier of hydrogen is water vapor. Therefore, we adopt $f_\mathrm{T} ( \mathrm{H}_2 ) = f( \mathrm{H}_2 \mathrm{O} )$, where $f( \mathrm{H}_2 \mathrm{O} )$ is the mixing ratio of water vapor in the upper atmosphere.

	An escape flux is limited by EUV flux if the planetary upper atmosphere is hydrogen-rich atmosphere. The energy-limited escape flux ($\phi_\mathrm{e}$ ) is given by \citet{Wat81} as
\begin{equation}
   \phi_\mathrm{e} = \frac{\epsilon S_\mathrm{EUV} r_\mathrm{p}}{m_\mathrm{i} G M_\mathrm{p} }
\end{equation}
where $\epsilon$ is the escape efficiency, $S_\mathrm{EUV}$ is the EUV radiation flux, $G$ is the gravitational constant,
$r_\mathrm{p}$ is the radius of the planet, and $M_\mathrm{p}$ is the planetary mass.
According to \citet{Wat81}, the escape efficiency ranges from $0.15$ to $0.3$; we adopted $0.15$ as the escape efficiency.
Since we used Earth-sized planets in this study, we used the Earth's radius and mass for $r_\mathrm{p}$ and $M_\mathrm{p}$, respectively.

	Atmospheric $\mathrm{H}_{2}\mathrm{O}$ molecules are dissociated by UV radiation from the central star (via $\mathrm{H}_{2}\mathrm{O} + \mathrm{h}\nu \rightarrow \mathrm{H} + \mathrm{OH}$) \citep[e.g.,][]{Bri69}. Hydrogen molecules are transported to the upper atmosphere, where EUV radiation reaches, and are heated by such radiation. The resultant hydrogen molecules escape to space\citep[e.g.,][]{Kas84}. When hydrogen escapes, the oxygen left behind can build up in the planetary atmosphere. Because we assume diffusion-limited escape, oxygen does not escape. If an amount of water equivalent to the Earth's ocean mass escapes, approximately $240 \, \mathrm{bars}$ of $\mathrm{O}_{2}$ will remain \citep[]{Kas97}. In this study, we assume that all of this oxygen is absorbed by the crust of the planet.

\subsection{The mixing ratio of water vapor in the upper atmosphere}

\begin{figure}
\epsscale{.80}
\plotone{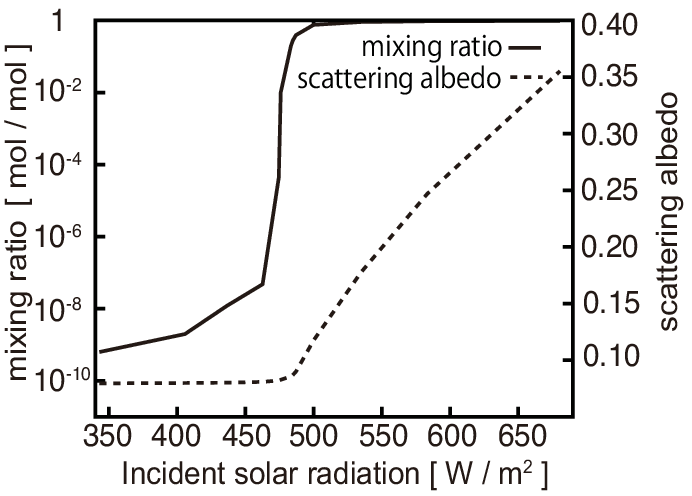}
\caption{
  The mixing ratio of water vapor in the upper atmosphere and scattering albedo by water vapor for an aqua planet as a function of the incident solar radiation. The mixing ratio of water vapor increases rapidly at around $470 \, \mathrm{W/m}^{2}$. In our model, we assume a constant albedo for the planetary surface ($0.3$) and the scattering albedo by water vapor, which varies with the amount of water vapor in the atmosphere.
  \label{Fig1}}
\end{figure}

	In order to estimate the escape flux of hydrogen molecules in the diffusion-limited escape mode, we need the mixing ratio of water vapor in the upper atmosphere. For the aqua planet mode, we estimate the mixing ratio of water vapor in the upper atmosphere as described by \citet{Abe88}. Figure \ref{Fig1} shows the mixing ratio of water vapor in the upper atmosphere as a function of the incident stellar radiation. It is very low until the cold trap region disappears. The escape flux of hydrogen molecules in the diffusion-limited escape mode is thus expected to be very low due to the very low mixing ratio of water vapor under these conditions. If an aqua planet receives incident solar flux from the central star that is sufficiently strong to eliminate the cold trap, the upper atmosphere will become humid and the escape of hydrogen molecules will be rapid.

\begin{figure}
\epsscale{.80}
\plotone{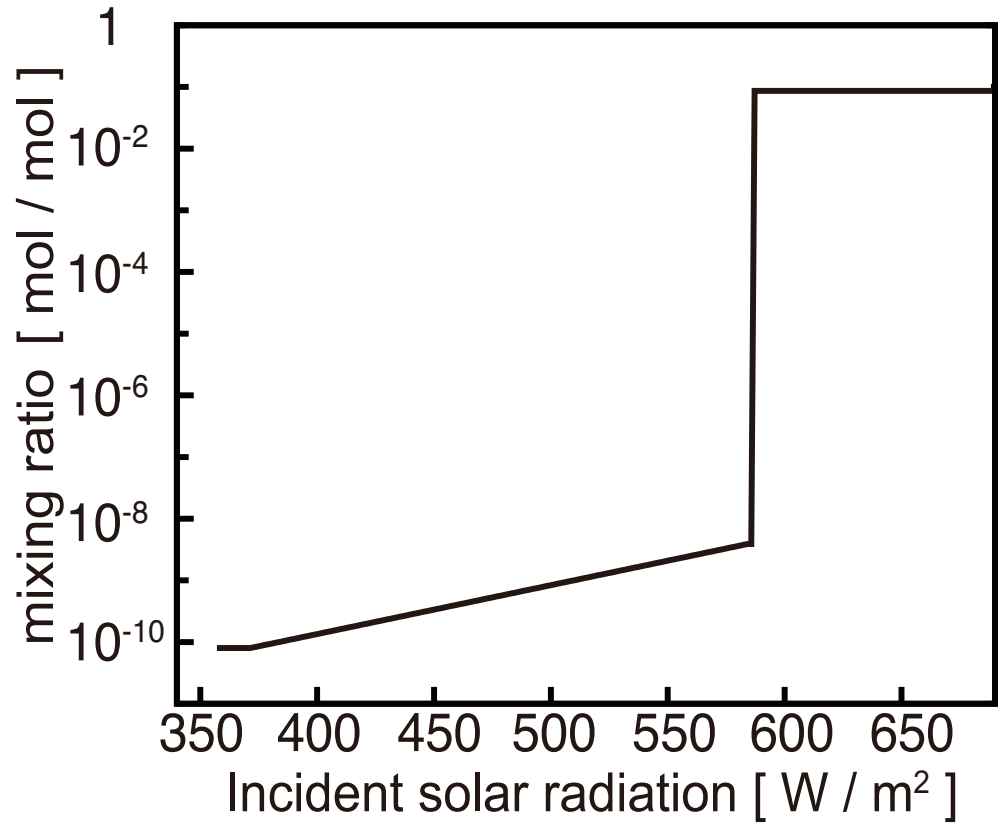}
\caption{
  The mixing ratio of water vapor in the atmosphere for a land planet as a function of the incident solar radiation. If a land planet receives the incident solar radiation exceeding the threshold of the runaway greenhouse ($585 \, \mathrm{W/m}^{2}$) for a land planet, then the mixing ratio of water vapor is high. In our model, we consider the constant albedo ($0.3$) throughout a period of time of a land planet because there is little water vapor before the threshold of the runaway greenhouse for a land planet.
  \label{Fig2}}
\end{figure}

	For the land planet mode, we estimate the mixing ratio of water vapor in the upper atmosphere using the results of \citet{Abe11}. The mixing ratio for a land planet also increases with increasing luminosity of the central star. In a case where the insolation from the central star is above the threshold for the runaway greenhouse effect for a land planet, the upper atmosphere becomes humid and the escape of hydrogen is rapid (Fig.\ref{Fig2}).

\subsection{Stellar evolution}

	We consider the evolution of total luminosity and the EUV flux of a star because the escape flux of hydrogen molecules depends strongly on both. The total luminosity of a star increases over time. We treated the central star as a solar type star. For the luminosity for a G-type star, we use the following expression given by \citet{Gou81}:
\begin{equation}
	S(t) = \left[ 1 + \frac{2}{5} \left( 1 - \frac{t}{t_{\odot} }  \right) \right]^{-1} S_{\odot}   \label{lumi}
\end{equation}
where $S(t)$ is the incident stellar flux for a G-type star at $1 \, \mathrm{AU}$ as a function of time,
$t$ is the time in Gyr, $t_{\odot}$ is the age of the Sun ( $t_{\odot} = 4.5 \, \mathrm{Gyr}$ ),
and $S_{\odot}$ is the solar constant ( $S_{\odot} = 1366 \, \mathrm{W/m}^{2}$ ).

	Since the EUV flux for young stars is stronger than that for old stars, we employed the following expression of \citet{Lam09} for a G-type star: 
\begin{equation}
	L_\mathrm{EUV} = \begin{cases}
			0.375 L_{0} t^{-0.425} & ( t \leq 0.6 \mathrm{Gyr} ) \\
			0.19 L_{0} t^{-1.69} & ( t > 0.6 \mathrm{Gyr} )
			\end{cases}
\end{equation}
where $L_\mathrm{EUV}$ is the luminosity of EUV in Watt s$^{-1}$, and $ L_{0}= 10^{22.35} \mathrm{W}$.
The relation between the luminosity and the flux is given by $S_{\mathrm{EUV} }= L_\mathrm{EUV} / {4 \pi a^{2}} $, where $a$ is a distance from the central star to the planet in meters.

%*************************************** Evolution model **********************************************

\section{Results}

\subsection{Effect of the initial amount of water}

\begin{figure}
%\epsscale{.80}
\plotone{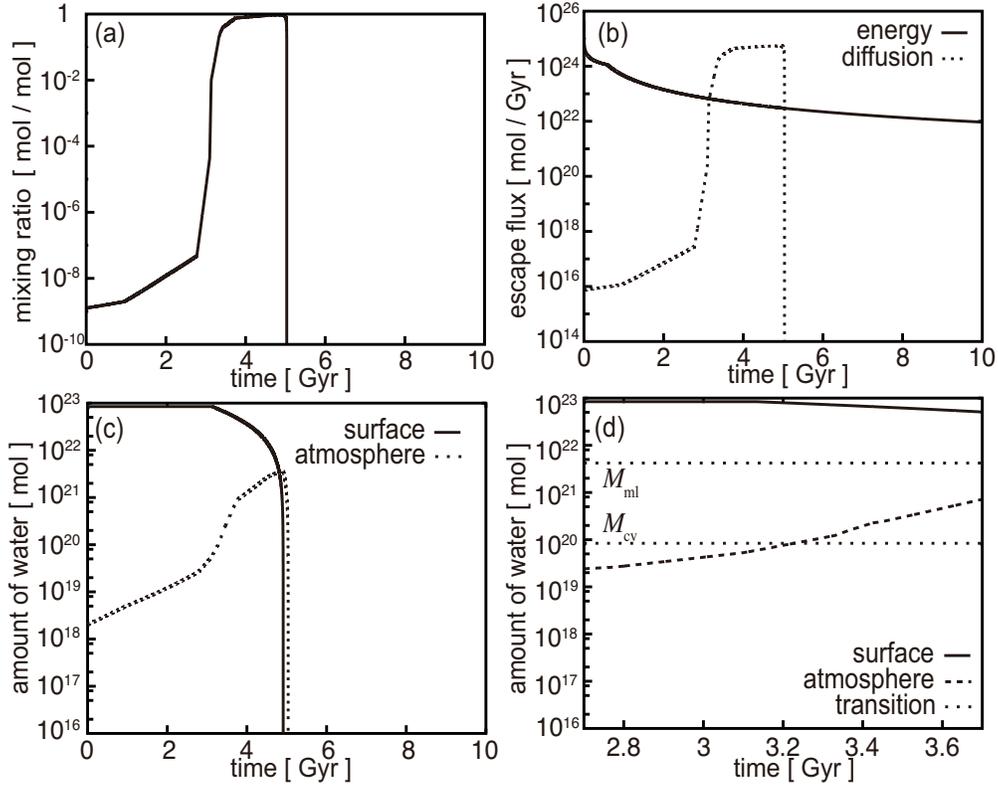}
\caption{Evolution of a planet with $M_\mathrm{ini} = 1.0 M_\mathrm{oc}$ at $a = 0.80$ AU. (a) The mixing ratio of water vapor in the upper atmosphere; (b) escape flux of hydrogen; (c and d) change in the amount of water in atmosphere and on the surface are shown as a function of time. Figure (d) shows the change in the amount of water around $3.2 \, \mathrm{Gyr}$ with two transition conditions (the critical amount of vapor for the runaway greenhouse effect, $M_\mathrm{cv}$, and the maximum liquid water mass for a land planet, $M_\mathrm{ml}$) from an aqua planet to a land planet. Here, $M_\mathrm{cv}$ 
are set a $3 \, \mathrm{m}$ in global average for the precipitable water and $M_\mathrm{ml}$ is set to $0.05 \, M_\mathrm{oc}$. In this case, the aqua planet evolves directly to a steam planet and then a dry planet; it is never a land planet.
    \label{Fig3}}
\end{figure}

	In this section, we examine typical results obtained by our numerical simulations describing the planetary evolution. Figure \ref{Fig3} shows the evolution of a planet with the present Earth's ocean 
at the beginning ($M_\mathrm{ini} = 1.0 \, M_\mathrm{oc}$) at $a = 0.80 \, \mathrm{AU}$, where $M_\mathrm{ini}$ is the initial amount of water.

	As long as the luminosity of the central star is low ($ < 3 \, \mathrm{Gyr}$), the mixing ratio of water vapor in the upper atmosphere remains low (Figure \ref{Fig3}a). Consequently, the escape flux of water vapor in the diffusion-limited escape mode also remains extremely low (Figure \ref{Fig3}b) and the amount of water on the surface hardly changes during this period (Figure \ref{Fig3}c). The atmosphere gets wetter as the luminosity of the central star increases over time, and after $3 \, \mathrm{Gyr}$, the mixing ratio of water vapor in the upper atmosphere increases rapidly. The increase in the mixing ratio in the upper atmosphere enhances the escape flux of water vapor in the diffusion-limited escape mode and decreases the amount of water on the planet (Figure \ref{Fig3}c). In this calculation, this planet does not satisfy the transition conditions from aqua planet mode to land planet mode because the amount of surface water exceeds the maximum water mass for land planet mode ($0.05 \, M_\mathrm{oc}$) when the average column mass of atmospheric water vapor reaches $M_\mathrm{cv}$ (i.e. a depth of $3 \, \mathrm{m}$), which is the critical amount for the runaway greenhouse effect. Therefore, the planet remains in aqua planet mode until the onset of the runaway greenhouse effect, eventually evolving into the uninhabitable steam planet state. Over a period of approximately $0.1 \, \mathrm{Gyr}$, all of the water escapes into space and the planet finally evolves into a dry planet state.

\begin{figure}
%\epsscale{.80}
\plotone{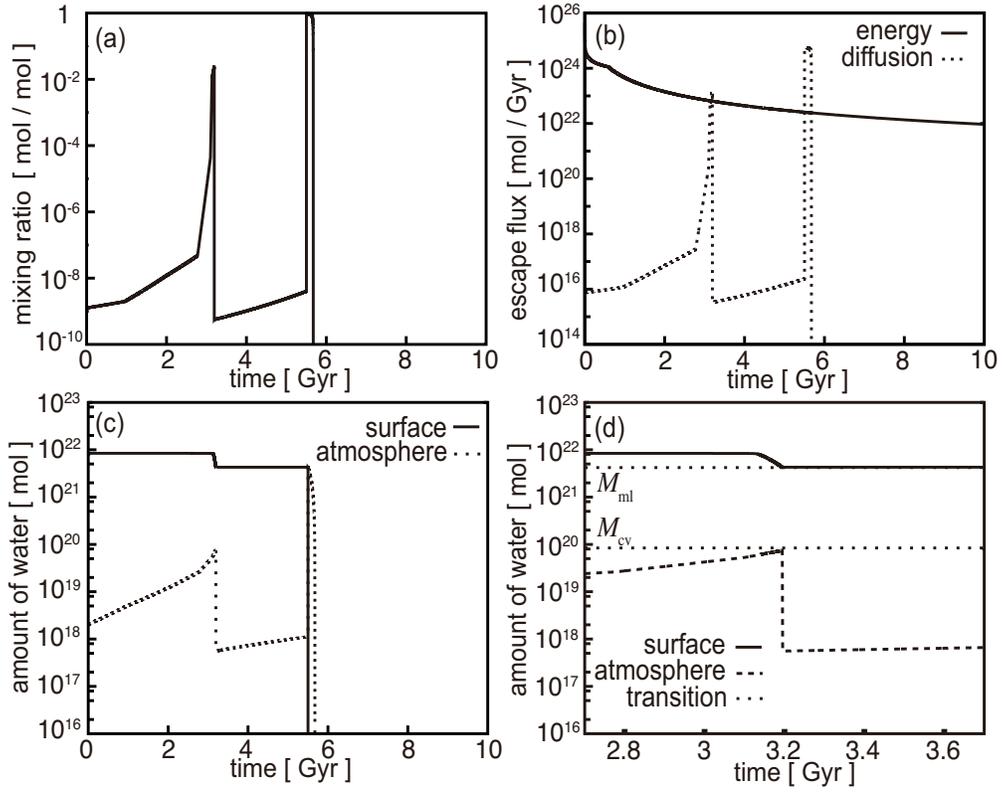}
\caption{Like Figure \ref{Fig3}, except that the planet has less initial water, $M_\mathrm{ini} = 0.1 \, M_\mathrm{oc}$ at $a = 0.80 \, \mathrm{AU}$. In this case, the aqua planet can evolve to a land planet and maintains liquid water on its surface for an additional $2 \, \mathrm{Gyrs}$.
    \label{Fig4}}
\end{figure}

	Figure \ref{Fig4} shows the evolution of a planet with $M_\mathrm{ini} = 0.1 \, M_\mathrm{oc}$ and $a = 0.8 \, \mathrm{AU}$. The difference between the cases shown in Figure \ref{Fig3} and Figure \ref{Fig4} is the initial amount of water. The overall evolution of water is very similar to that shown in Figure \ref{Fig3} before $t = 3.19 \, \mathrm{Gyr}$. However, when the amount of surface water decreases to $M_\mathrm{ml} = 0.05 \, M_\mathrm{oc}$ (i.e. the maximum water mass for the land planet mode), the average column mass of the atmospheric water vapor is less than the $3 \, \mathrm{m}$ threshold for the runaway greenhouse effect in figure \ref{Fig4}d. This planet therefore satisfies the requirements for a transition from aqua planet mode to land planet mode and can evolve into land planet mode. Once in land planet mode, the planet's atmosphere becomes dry (Figure \ref{Fig4}a) and escape shuts off (Figure \ref{Fig4}b). This planet can keep liquid water on its surface until the central star has become sufficiently bright enough to exceed the runaway threshold of a land planet (Figure \ref{Fig4}c). When the planet receives insolation above this limit, the planet evolves into the steam planet state and then the dry planet state once it has lost its remaining water into space over about $0.17 \, \mathrm{Gyr}$.

	Whether or not an aqua planet can evolve to a land planet thus depends on the initial amount of water. Qualitatively, planets with small initial amounts of water can more easily evolve from the aqua planet mode to the land planet mode.

\subsection{Effect of the distance from the central star}

\begin{figure}
%\epsscale{.80}
\plotone{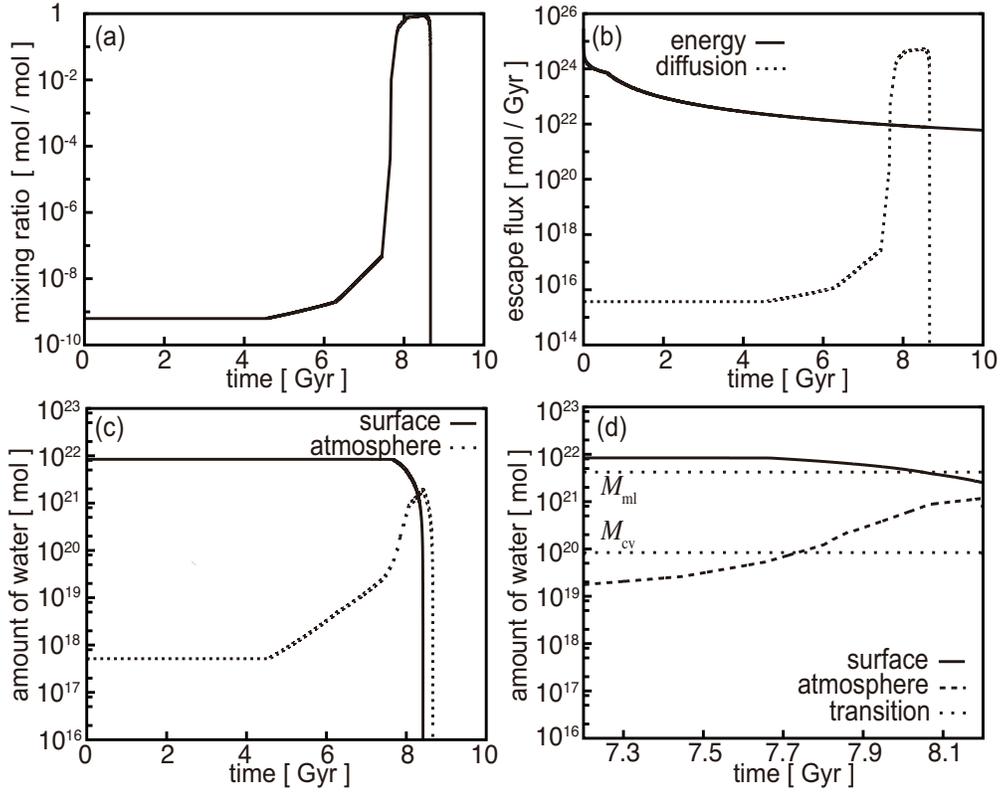}
\caption{Like Figure \ref{Fig4}, but for a planet a little farther from the central star, with $M_\mathrm{ini} = 0.1 \, M_\mathrm{oc}$ at $a = 1.0 \, \mathrm{AU}$. Figure (d) shows the change in the amount of water at approximately $7.7 \, \mathrm{Gyr}$. This aqua planet does not pass through a stage as a land planet, but instead evolves directly into a steam planet and then a dry planet.
    \label{Fig5}}
\end{figure}

	Here we examine the effect of the distance from the central star on planetary evolution. Figure \ref{Fig5} shows the evolution of a planet with $M_\mathrm{ini} = 0.1 \, M_\mathrm{oc}$ and $a = 1.0 \, \mathrm{AU}$. In Figure \ref{Fig5}d, the amount of surface water remaining on the planet exceeds $M_\mathrm{ml}$ and the planet experiences the runaway greenhouse effect. This planet evolves into the steam planet state without passing through a land planet state.  On the other hand, we have seen that a planet with the same initial amount of water at $0.8 \, \mathrm{AU}$ can evolve from aqua planet mode to land planet mode (Figure \ref{Fig4}). This evolution occurs because the planet nearer the central star evolves through the moist greenhouse stage when the star is younger, and the EUV flux and hydrogen escape flux are higher. Any planet near the central star will lose more water during the period of rapid escape of water vapor.

	The timing of the onset and duration of rapid escape of water vapor are the two keys that determine whether or not an aqua planet will evolve into a land planet. In the case of a planet that is located far away from its central star, the onset of rapid escape of water vapor occurs later when the star is older and its EUV flux is lower. It therefore takes more time for a more distant planet to lose its surface water; phrased another way, the more distant planet loses less water between the onset of the moist greenhouse effect and the onset of the runaway greenhouse effect for an aqua planet. Moreover, because the rate of stellar luminosity increases with time (see (\ref{lumi})), more distant planets have less time to shed excess water before the onset of runaway greenhouse effect. As a result, it is harder for a distant planet to evolve into a land planet, and easier to evolve directly into a steam planet.

\subsection{Limit of the initial amount of water to evolve to a land planet} \label{limit}

	In the previous section, we showed several typical numerical results obtained for planetary evolution. We found that the initial amount of water ($M_\mathrm{ini}$) and the distance from the central star ($a$) are important in determining whether or not an aqua planet can evolve into a land planet.

\begin{figure}
%\epsscale{.80}
\plotone{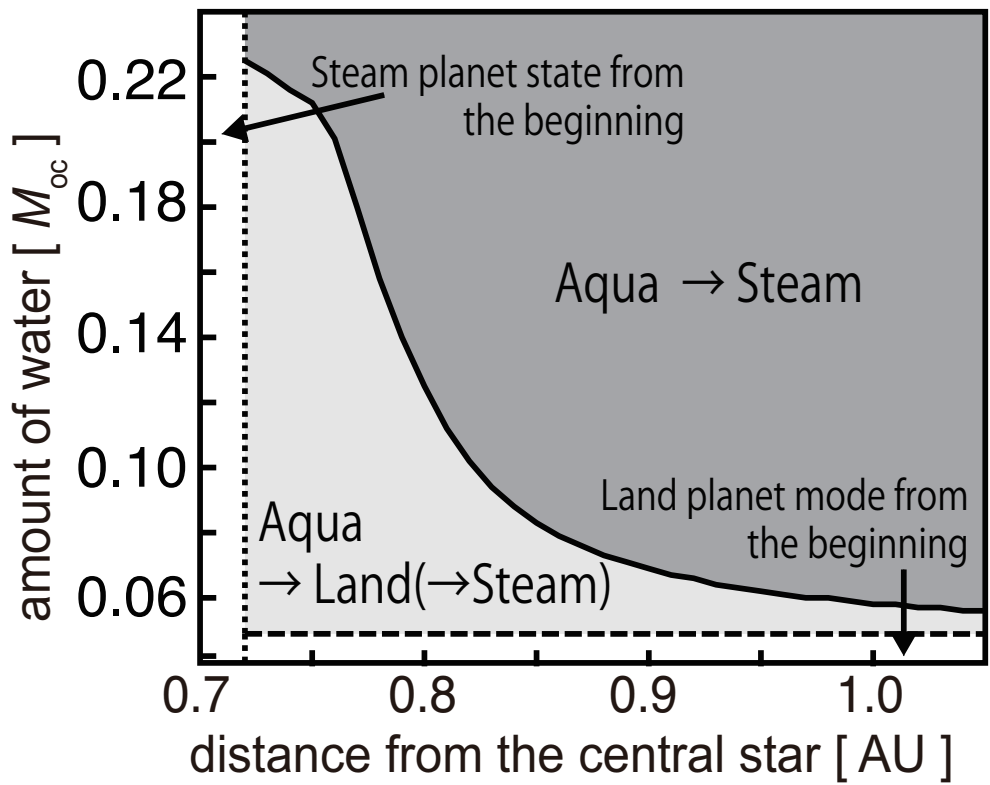}
\caption{Initial planetary conditions required for the evolution of a land planet with a central star like the Sun. Planets at $a < 0.72 \, \mathrm{AU}$ are Type II planets \citep[]{Ham13}, which are in steam planet state in the beginning and lose their water before cool down. Thus, they never have liquid water on their surface. On the other hand, planets at $a > 0.72 \, \mathrm{AU}$ and with $M_\mathrm{ini} < 0.05 \, M_\mathrm{oc}$ may be land planets from the beginning.
    \label{Fig6}}
\end{figure}

	Figure \ref{Fig6} shows the maximum initial amount of water that is required for the transition from the aqua planet mode to the land planet mode as a function of the distance from the central star. In this figure, planets below the curve can evolve from aqua planet mode to land planet mode. Aqua planets close to a central star with small initial amounts of water can evolve into land planets, but aqua planets that are too far from a central star, or that are born with large amounts of water, evolve into steam planets without passing through the land planet stage.

	Planets with less than the maximum liquid water mass for land planets ($M_\mathrm{ini} < M_\mathrm{ml}$) remain in land planet mode from the beginning. In cases of planets with small orbits ($a < 0.72 \, \mathrm{AU}$), all of the water evaporates and such planets are in a steam planet state from the beginning.

\subsection{Inner edge of the HZ}

\begin{figure}
%\epsscale{.80}
\plotone{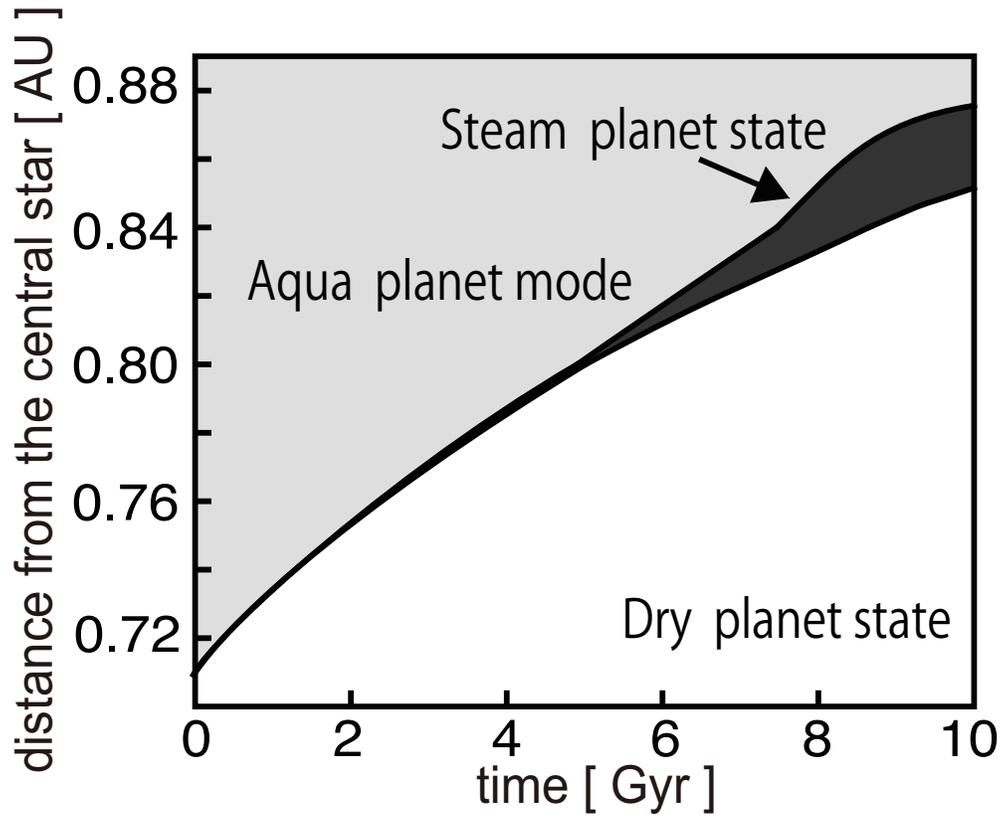}
\caption{The evolution of planets with $M_\mathrm{ini} = 1.0 \, M_\mathrm{oc}$ for central stars like the Sun. None of these planets become land planets. They have too much water to start with. Note that some of the planets that orbit older stars spend considerable time as steam planets.
    \label{Fig7}}
\end{figure}

	Here we discuss the inner edge of the HZ of solar-type stars based on our results. Figure \ref{Fig7} shows the evolution of planets with $M_\mathrm{ini} = 1.0 \, M_\mathrm{oc}$ at various distances from the central star under the same transition conditions. For planets with $M_\mathrm{ini} = 1.0 \, M_\mathrm{oc}$, rapid escape of water occurs, but such planets cannot evolve to the land planet mode because they cannot lose enough water -- they cannot get from $1.0 \, M_\mathrm{oc}$ to $M_\mathrm{ml}$ -- before encountering the runaway greenhouse threshold for aqua planets. Therefore, the boundary between the aqua planet mode and the steam planet state corresponds to the inner edge of the classical instantaneous habitable zone, proposed in \citet{Kas93} (Figure \ref{Fig7}).

\begin{figure}
%\epsscale{.80}
\plotone{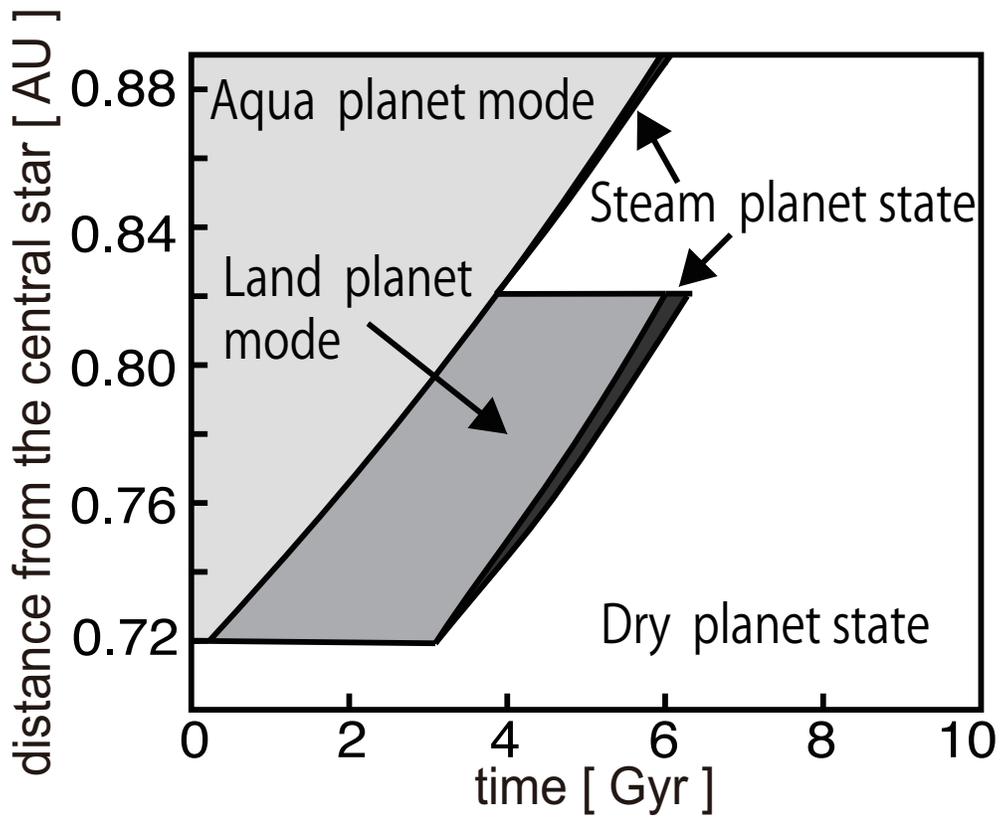}
\caption{Like Figure \ref{Fig7} for initially drier planets with $M_\mathrm{ini} = 0.1 \, M_\mathrm{oc}$. Aqua planets between $0.72$ and $0.82 \, \mathrm{AU}$ can transform into land planets, and as land planets they maintain liquid water on their surfaces for an additional $2 \, \mathrm{Gyrs}$. Thereafter, such planets receive the solar radiation of the threshold of the runaway greenhouse for a land planet and become steam planets.
    \label{Fig8}}
\end{figure}

	As in Figure \ref{Fig7}, Figure \ref{Fig8} shows the time evolution of planets with $M _\mathrm{ini} = 0.1 \, M_\mathrm{oc}$. When located between $0.72$ and $0.82 \mathrm{AU}$, such planets can evolve from aqua planet mode to land planet mode. Planets in the land planet mode remain habitable for an additional $2 \, \mathrm{Gyrs}$ before evolving into the steam planet state and then to the dry planet state due to increased luminosity of the central star.

In both Figure \ref{Fig7} and Figure \ref{Fig8}, aqua planets closer than about $0.7 \, \mathrm{AU}$ from the central star are in the steam planet state from the beginning. Such planets lose all of their water rapidly (over a few tens of million years) because the EUV flux at that time is very strong; such planets are called Type II terrestrial planets \citep[]{Ham13}.

	Figure \ref{Fig9} shows the inner edge of the CHZ, which is defined here as the region where a planet has liquid water on its surface for more than $4.6 \, \mathrm{Gyr}$. We divided the CHZ into two types according to the initial amount of water. One type of CHZ refers to those regions on aqua planets that maintain water on their surface for more than $4.6 \, \mathrm{Gyr}$; this type corresponds to the CHZ proposed in \citet{Kas93} and is referred to here as a type-I CHZ. The other type of CHZ refers to regions on land planets that maintain water on their surface for a total of more than $4.6 \, \mathrm{Gyr}$; this type is referred to as a type-II CHZ.

	In our model, planets with $ M_\mathrm{ini} > 0.22 \, M_\mathrm{oc}$ cannot evolve to land planets: such planets have a type-I CHZ. On the other hand, planets with $M_\mathrm{ini} < 0.22 \, M_\mathrm{oc}$ can evolve into land planets. However, planets with $ 0.19 \, M_\mathrm{oc} < M_\mathrm{ini} < 0.22 \, M_\mathrm{oc}$ cannot maintain liquid water on their surface for $4.6 \, \mathrm{Gyrs}$ because the insolation on such planets exceeds the threshold of the runaway greenhouse effect. Planets with $M_\mathrm{ini} \leq 0.19 \, M_\mathrm{oc}$ have a type-II CHZ.

	Figure \ref{Fig9} shows that planets that begin with $0.09 \, M _\mathrm{oc} < M_\mathrm{ini} \leq 0.19 \, M_\mathrm{oc}$ have both type-I and type-II CHZs, but these two CHZs are not contiguous. However, planets with $0.05 \, M_\mathrm{oc} < M_\mathrm{ini} \leq 0.09 \, M_\mathrm{oc}$  also have both CHZs and they are contiguous. Planets with an initial water less than the maximum water mass for a land planet ($M_\mathrm{ml} = 0.05 \, M_\mathrm{oc}$) on their surface are in land planet mode from the beginning.

%*************************************** Results **********************************************

%*************************************** Discussion **********************************************

\section{Discussion}

\subsection{Effect on the transition condition}

	In previous sections, we showed results when $M_\mathrm{cv}$, the critical water vapor amount for the runaway greenhouse, was taken as a global average of $3 \, \mathrm{m}$ water depth equivalent and $M_\mathrm{ml}$, the maximum liquid water mass for a land planet mode, was taken to be $0.05 \, M_\mathrm{oc}$. In this section, we investigate how the evolution from aqua planet mode to land planet mode changes when the transition conditions  are altered.

	Figure \ref{Fig10} shows the maximum $M_\mathrm{ini}$ required for a planet to evolve into a land planet for different values of $M_\mathrm{cv}$. In these calculations, $M_\mathrm{ml}$ is fixed at $0.05 \, M_\mathrm{oc}$. If $M_\mathrm{cv}$ exceeds $3 \, \mathrm{m}$, then the maximum $M_\mathrm{ini}$ increases because the onset of the runaway greenhouse effect is delayed. For planets closer to the central star, the maximum $M_\mathrm{ini}$ is larger because the escape flux of water is larger before the onset of runaway greenhouse effect. However, if planets are too close to the central star, the maximum $M_\mathrm{ini}$ decreases and has a peak (see Figure \ref{Fig10}). The amount of water that escapes is determined by the duration of the period of a rapid escape until the onset of the runaway greenhouse effect and the escape flux of water during this period. For planets that are very close to the central star, runaway greenhouse effect occurs early and duration of the period of rapid escape is short. Although the escape flux in the energy-limited escape mode is large, the rapid escape of water occurs in the diffusion-limited escape mode for these planets. Therefore, the total amount of escaped water hits a peak, which is shown by the maximum $M_\mathrm{ini}$ value in Figure \ref{Fig10}. An aqua planet cannot evolve into a land planet when $M_\mathrm{cv}$ is $1 \, \mathrm{m}$ (global average). In our model, the mixing ratio of water vapor in the upper atmosphere reached $10^{-3}$ when the column of water in the atmosphere is $1.9 \, \mathrm{m}$ (global average). If $M_\mathrm{cv}$ is less than $1.9 \, \mathrm{m}$, then a planet in the aqua planet mode cannot lose enough water and it passes directly into the steam planet state.

	Our standard case takes $M_\mathrm{ml} = 0.05 \, M_\mathrm{oc}$. This value of $M_\mathrm{ml}$ is dependent upon the planetary topography.  Figure \ref{Fig11} shows the dependence of the maximum $M_\mathrm{ini}$ on $M_\mathrm{ml}$ for a planet evolving into a land planet. In this calculation, the value of $M_\mathrm{cv}$ is fixed at our standard value of $3 \, \mathrm{m}$. A smaller value for $M_\mathrm{ml}$ results in smaller maximum $M_\mathrm{ini}$, but it is possible for aqua planets to evolve into land planets even if $M_\mathrm{ml} = 0.01 \, M_\mathrm{oc}$. However, an aqua planet cannot evolve into a land planet when $M_\mathrm{cv}$ is less than $1 \, \mathrm{m}$ (see Figure \ref{Fig10}). Therefore, the parameter $M_\mathrm{cv}$ has more impact on the evolution path to land planets than the parameter $M_\mathrm{ml}$.

\subsection{Atmospheric models}

	We investigated the evolutionary paths of planets using a $1$-D, cloud-free, radiative-convective atmospheric model developed in the $1980$s \citep[]{Abe88}. The $1$-D atmospheric models have been improved by using more accurate opacity estimates for water vapor \citep[]{Kop13, Gol13, Kop14}. The inner edge of the CHZ (defined by moist greenhouse effect) in \citet{Kop13} is $0.99 \, \mathrm{AU}$. In our model, the onset of the moist greenhouse effect is $1.39 \, S_{\odot}$. Since the onset of the runaway greenhouse effect is dependent on the value of $M_\mathrm{cv}$, when $M_\mathrm{cv}$ is set to $3 \, \mathrm{m}$, the onset of the runaway greenhouse effect is $1.40 \, S_{\odot}$. In \citet{Kop13}, the onsets of the moist and runaway greenhouse effects are $1.015$ and $1.06 \, S_\odot$, respectively.  The timing of the onset of rapid water loss and the amount of water that escaped should therefore be different between our atmospheric model and their model. In order to investigate the evolution from an aqua planet to a land planet, we need the relationship between the incident solar flux and the amount of water vapor in the atmosphere, but it was not shown in the previous papers. We therefore qualitatively compared our results to those of \citet{Kop13}.

	We estimated how much water escapes during the period from the onset of the moist greenhouse effect to the onset of the runaway greenhouse effect using the results from \citet{Kop13}. To estimate the escaped water amount by \citet{Kop13}, we presume that the mixing ratio of water vapor is $10^{-3}$, so that the escaped water amount during this period in the diffusion-limited escape mode is approximately $3 \times 10^{21} \, \mathrm{moles}$ (ca. $0.04 \, M_\mathrm{oc}$). The amount of escaped water under energy-limited escape mode conditions in the same period was estimated to be $10^{23} \, \mathrm{moles}$ (ca. $1.19 \, M_\mathrm{oc}$). The actual amount of escaped water falls between these estimates (i.e., between $ 3 \times 10^{21}$ and $1 \times10^{23} \mathrm{moles}$).  In our model, planets that can evolve into land planets typically lose about $0.1 \, M_\mathrm{oc}$ of water (see Section \ref{limit}), which is comparable to, or less than, the estimates of \citet{Kop13}. \citet{Kop13} indicates that planets at $a < 0.84 \, \mathrm{AU}$ are in the steam planet state. Thus, the region where aqua planets can evolve into land planets moves outward (about $0.1$ AU) when the model of \citet{Kop13} is applied. The type-II CHZ (see Figure \ref{Fig9}), in which aqua planets evolve to land planets, also moves outward.

\subsection{Intensity of EUV flux}

	We assessed the effect of EUV flux on our results. The uncertainty associated with the EUV flux from the central star is large \citep[e.g.,][]{Zah82, Lam09}. Just before an aqua planet evolves into a land planet, water escapes rapidly in the energy-limited escape mode (see Figure \ref{Fig4}b). If the EUV flux is larger than that used in our model, then the escape flux increases. However, since the escape flux is limited by the diffusion process, a larger EUV flux does not markedly affect our results. On the other hand, if the EUV flux is smaller than that used in our model, then the escape flux decreases. This makes it harder for aqua planets to evolve into land planets.

\subsection{Amount of atmospheric $\mathrm{CO}_2$}

	An important complication could be the evolution of the amount of $\mathrm{CO}_2$ in atmosphere. Since the carbon cycle on a land planet has not been investigated yet, we speculate possible effects of carbon cycle on a land planet climate in the following.

	The amount of $\mathrm{CO}_2$ in the atmosphere is governed by the balance between the degassing flux of $\mathrm{CO}_2$ and the removal flux of $\mathrm{CO}_2$ through chemical weathering. We expect the degassing flux of $\mathrm{CO}_2$ to be smaller on a land planet than on an aqua planet because plate tectonics is likely to be less efficient on a water-poor mantle. Chemical weathering may be also less efficient on a land planet, because it occurs mainly in wet regions. If the former effect exceeds the latter, then the amount of $\mathrm{CO}_2$ in the atmosphere should be smaller on a land planet than on an aqua planet. However, any decrease in the degassing flux would be difficult to estimate.

	Therefore, it is not clear whether there would be more or less $\mathrm{CO}_2$ on a land planet compared to an aqua planet. If the amount of $\mathrm{CO}_2$ is balanced at a high value, then the period of habitability may be shorter than that estimated in our study because the greenhouse effect attributed to $\mathrm{CO}_2$ may trigger the runaway greenhouse effect. On the other hand, if the $\mathrm{CO}_2$ amount is balanced at a low value, the environment is cooler than our estimate. However, such an effect is likely minor, because the insolation is high enough to keep from the global freezing without $\mathrm{CO}_2$ greenhouse effect. It should also be noted that a land planet is relatively resistant to global freezing\citep[]{Abe11}.  Even if ice caps appear, their size is regulated by the carbon cycle, because $\mathrm{CO}_2$ removal through chemical weathering would be negligible in the ice-covered area. Under such a scenario, the lifetime of a habitable environment would likely be similar to that estimated in our study.

\subsection{The geologic H$_2$O cycle}

We have neglected the geologic water cycle by presuming that the fluxes of water into and out of the mantle are in rough balance. 
This rough balance is likely to break down after an aqua planet has become a land planet, because the mantle may still be wet and outgassing may still be considerable, but subduction of hydrous minerals is probably restricted to locations near the poles.  Thus water outgassing would be a potential hazard to the future habitability of the land planet over long time scales.  But while the aqua planet is losing its hydrogen in the moist greenhouse state, it is possible that the imbalance goes in the other directions, as the generally warm wet climate of the moist greenhouse would likely promote weathering reactions and possibly promote subduction of hydrous minerals.  Thus we think it probable that the geologic water cycle would aid rather than subvert the transition of a suitable planet from aqua to land modes.

\subsection{Other stars}

Stars of different spectral types evolve differently from the Sun. 
In particular, the luminosity evolution of stars of later spectral type (late G dwarfs, K dwarfs, and M dwarfs) is slower
than in the Sun.  These stars are fainter than the Sun and hence the HZ is closer to these stars than to the Sun.
The ratio of EUV radiation to luminosity is about the same for K stars as for G stars,
and the ratio is higher for most M stars \citep[]{Lam09}.
Hence these stars are at least as effective at driving hydrogen escape as a Sun-like star,
and so we would expect similar or faster rates of hydrogen escape for an HZ planet around these other stars. 
The slower rate of luminosity evolution means that there is more time available for escape to take place before the moist greenhouse becomes a runaway greenhouse,
and therefore we expect that late type stars are more favorable to a planet making a continuously habitable transition from aqua state to land state. We will quantitatively investigate the evolution path from the aqua planet mode to the land planet mode for stars of different spectral types in the next paper.

%*************************************** Discussion **********************************************

%*************************************** Conclusion **********************************************

\section{Conclusion}

	Planets can evolve from aqua planet mode to land planet mode and, in so doing, maintain liquid water on their surface for a longer time. Here we assess the various requirements of such a transition. We argue that the transition from an aqua planet to a land planet can occur if the amount of water on the surface is reduced rapidly by hydrogen escape via the photodissociation of $\mathrm{H}_{2}\mathrm{O}$ vapor, while the amount of water vapor in the atmosphere remains sufficiently low to prevent the onset of runaway greenhouse effect; i.e., the planet must lose most of its water while in the moist greenhouse state. We considered two factors as being central to the evolution of a land planet from an aqua planet: the critical vapor amount for the runaway greenhouse, $M_\mathrm{cv}$, and the maximum liquid water mass for a land planet mode, $M_\mathrm{ml}$. We assumed that the star in our simulation was similar to the Sun. We examined temporal changes in the amount of water on a planet under conditions of different initial amounts of water and orbital distance.  We illustrated the conditions for the evolution of an aqua planet to a land planet and applied this information to re-evaluate the inner edge of the HZ.

	The evolutionary paths of aqua planets are mostly determined by the initial amount of water, $M_\mathrm{ini}$. Planets with large $M_\mathrm{ini}$ evolve into the runaway greenhouse state before entering the land planet mode, while planets with small $M_\mathrm{ini}$ can evolve into land planets. Planets nearer the central star can lose water more quickly because these events take place when the star is younger and thus a stronger source of the EUV radiation that drives hydrogen escape. These findings imply that it is generally easier for a land planet to form when it is closer to its star than when it is far from its star.

	If $M_\mathrm{cv}$ is equivalent to $3 \, \mathrm{m}$ of precipitable water, then an aqua planet with $M_\mathrm{ini} = 0.1 \, M_\mathrm{oc}$ at $0.8 \, \mathrm{AU}$ can evolve into a land planet. In addition, such a planet can retain liquid water on its surface for an additional $2$ Gyr as a land planet. On the other hand, if $M_\mathrm{cv}$ is less than about $1 \, \mathrm{m}$, it is very difficult for any aqua planet to evolve into a land planet. This value for $M_\mathrm{cv}$ corresponds to the amount of water vapor in the atmosphere that causes the rapid water loss of a substantial fraction of the Earth's ocean mass. By contrast, the dependence of the evolutionary path on $M_\mathrm{ml}$ (the upper bound on the amount of water that defines a land planet) is not very large, implying that the value of $M_\mathrm{cv}$ is more important as a transition condition from an aqua planet to a land planet.

\begin{figure}
%\epsscale{.80}
\plotone{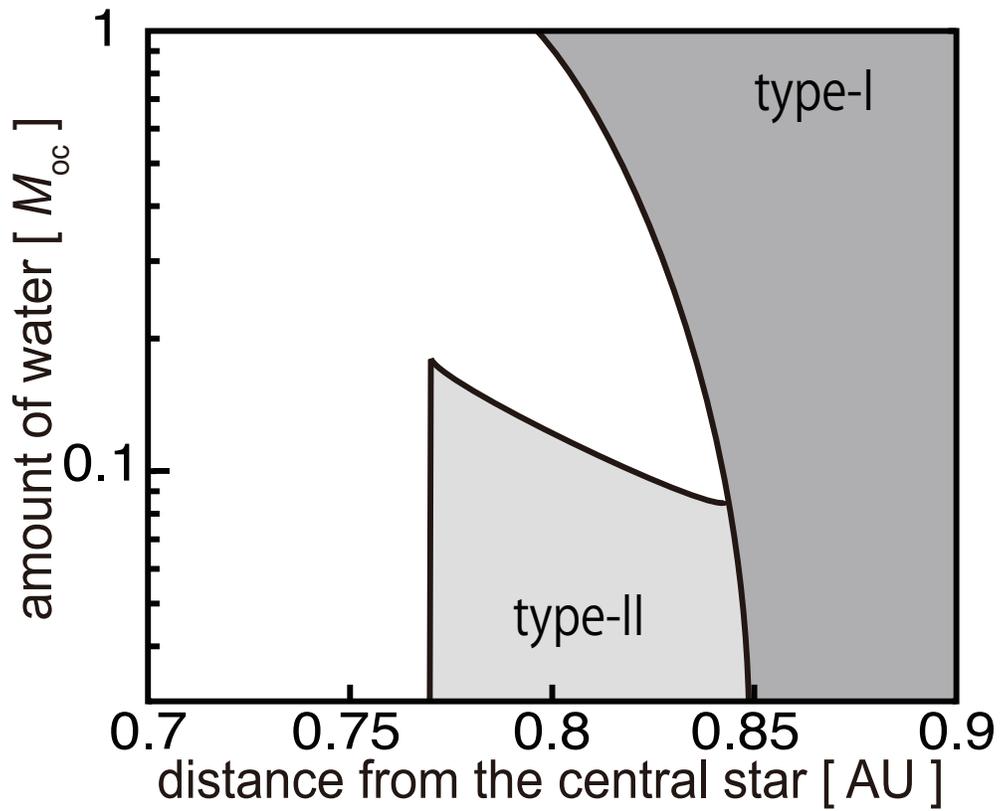}
\caption{The inner edge of the CHZ, defined as the planet being in the HZ for at least $4.6 \, \mathrm{Gyr}$, considering the evolution from an aqua planet to a land planet. The region of the type-I CHZ corresponds to the traditional CHZ for an aqua planet, while the region of the type-II CHZ is a region where initially aqua planets become land planets.
    \label{Fig9}}
\end{figure}

\begin{figure}
%\epsscale{.80}
\plotone{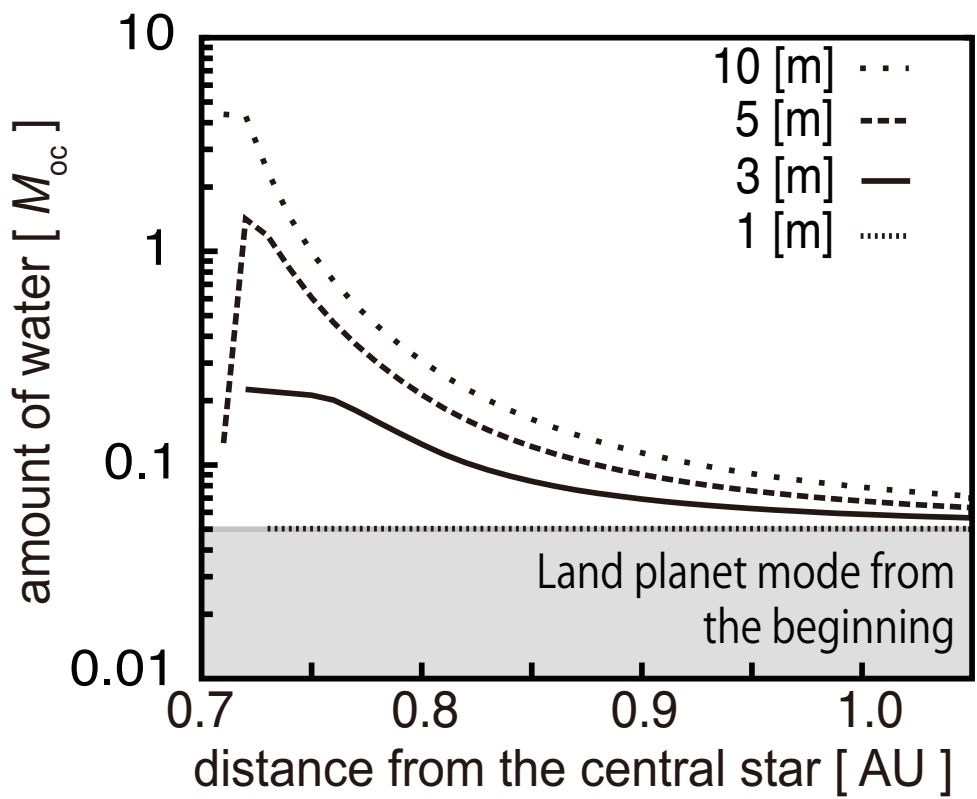}
\caption{The effect of the critical vapor amount for the runaway greenhouse $M_\mathrm{cv}$ on the maximum initial amount of water, $M_\mathrm{ini}$ to evolve from the aqua planet mode to the land planet mode. The maximum water mass for a land planet, which is the other transition condition, is fixed at $M_\mathrm{ml} = 0.05 \, M_\mathrm{oc}$. The result for (solid curve) the case where $M_\mathrm{cv}$ is equivalent to a column of $3 \, \mathrm{m}$ of precipitable water is the same as that shown in Figure \ref{Fig6}. Larger critical  amount of vapor facilitates the evolution to land planets. However, if $M_\mathrm{cv}$ is less than $1 \, \mathrm{m}$, it is very difficult for an aqua planet to evolve into a land planet.
    \label{Fig10}}
\end{figure}

\begin{figure}
%\epsscale{.80}
\plotone{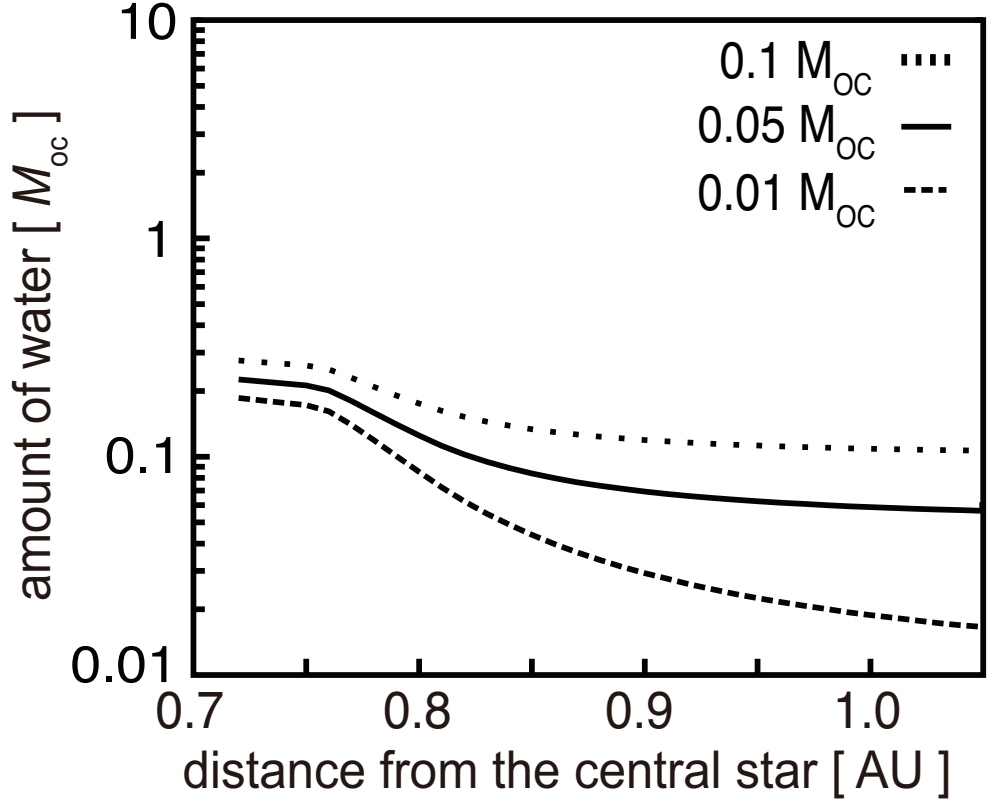}
\caption{The effect of the maximum water mass for a land planet, $M_\mathrm{ml}$, on the maximum initial amount of water, $M_\mathrm{ini}$ to evolve from the aqua planet mode to the land planet mode. Here $M_\mathrm{cv}$, the other transition condition, is fixed to $3$ m depth in global average. The result for the case of $M_\mathrm{ml} = 0.05 \, M_\mathrm{oc}$ was shown in Figure \ref{Fig6}. Changing $M_\mathrm{ml}$ has a smaller effect on the upper limit of initial water mass to evolve from an aqua planet to a land planet than changing $M_\mathrm{cv}$. The amount of water for a land planet from the beginning is corresponded with the maximum water amount for a land planet.
    \label{Fig11}}
\end{figure}

	We describe a new type of continuous habitable zone (CHZ) where aqua planets can evolve into land planets and maintain liquid water on their surface for more than $4.6 \, \mathrm{Gyr}$(type-II CHZ in Figure \ref{Fig9}). Rapid escape of water is not necessarily the end of a planet's habitability; rather, such water loss can bring about a new and different kind of habitability.

\acknowledgments

This work was supported by Grant-in-Aid for Scientific Research on Innovative Areas from the Ministry of Education, Culture, Sports, Science and Technology (MEXT) (No. 23103003).
This work was also partly supported by GCOE program "From Earth to Earths" (MEXT) and research Grant 2015 of Kurita Water and Environment Foundation.

\clearpage

%% the file, uncomment the \plotone, \plottwo, and
%% If you need a layout that cannot be achieved with \plotone or
%% \plottwo, you can invoke the graphicx package directly with the
%% \includegraphics command or use \plotfiddle. For more information,
%% please see the tutorial on "Using Electronic Art with AASTeX" in the
%% documentation section at the AASTeX Web site, http://aastex.aas.org/
%%
%% The examples below also include sample markup for submission of
%% supplemental electronic materials. As always, be sure to check
%% the instructions to authors for the journal you are submitting to
%% for specific submissions guidelines as they vary from

%% This example uses \plotone to include an EPS file scaled to
%% 80% of its natural size with \epsscale. Its caption
%% has been written to indicate that additional figure parts will be
%% available in the electronic journal.

\end{document}